\documentclass{aa}  

\usepackage{graphicx}
\usepackage{txfonts}
\usepackage[unicode=true,colorlinks=true,linkcolor=red,citecolor=blue,filecolor=green,urlcolor=green]{hyperref}

\usepackage{rotating}
\usepackage{lscape}
\usepackage{adjustbox}
\usepackage[toc,page]{appendix}

\newcommand{\nustar}{\textit{NuSTAR}}

\newcommand{\bjet}{\texttt{bjet}}
\newcommand{\dbb}{\texttt{diskbb}}
\newcommand{\xcp}{\texttt{xillverCp}}
\newcommand{\rlp}{\texttt{relxilllpCp}}
\newcommand{\zsh}{\texttt{zashift}}
\newcommand{\cons}{\texttt{constant}}
\newcommand{\tba}{\texttt{tbabs}}
\newcommand{\xspec}{\textrm{XSPEC}}

\begin{document}

   \title{A NuSTAR view of SS433}

   \subtitle{Precessional evolution of the jet--disk system}

   \author{F. A. Fogantini\inst{1,2}
          \and
          F. García\inst{1,2}
          \and
          J. A. Combi\inst{1,2,3}
          \and
          S. Chaty\inst{4}
          \and
          J. Martí\inst{5}
          \and 
          P. L. Luque~Escamilla\inst{3}
          }

   \institute{
    Instituto Argentino de Radioastronomía (CONICET; CICPBA), C.C. No 5, 1894 Villa Elisa, Argentina
    \and
    Facultad de Ciencias Astronómicas y Geofísicas, Universidad Nacional de La Plata, Paseo del Bosque, B1900FWA La Plata, Argentina
    \and
    Departamento de Ingeniería Mecánica y Minera (EPSJ), Universidad de Jaén, Campus Las Lagunillas s/n Ed. A3, E-23071 Jaén, Spain
    \and
    Université Paris Cité, CNRS, AstroParticule et Cosmologie, F-75013 Paris, France
    \and
    Departamento de F\'\i sica (EPSJ), Universidad de Ja\'en, Campus Las Lagunillas s/n, A3, 23071 Ja\'en, Spain \\
    }


  \abstract
   {SS433 is a Galactic microquasar with powerful outflows (double jet, accretion disk and winds) with well known orbital, precessional and nutational period. }
   {In this work we characterise different outflow parameters throughout the precessional cycle of the system.}
   {We analyse 10 NuSTAR (3--70 keV) observations of $\sim$30~ks that span $\sim$1.5 precessional cycles. We extract averaged spectra and model them using a combination of a double thermal jet model (\bjet) and pure neutral and relativistic reflection (\xcp\ and \rlp) over an accretion disk.}
   {We find an average jet bulk velocity of $\beta = v/c \sim0.29$ with an opening angle of $\lesssim$6~degrees. Eastern jet kinetic power ranges from 1 to $10^{39}$~erg/s, with base "coronal" temperatures $T_o$ ranging between 14 and 18 keV. Nickel  to iron abundances remain constant at $\sim$9 (within 1$\sigma$). The western to eastern jet flux ratio becomes $\sim1$ on intermediate phases, about 35\% of the total precessional orbit.
   The 3--70 keV total unabsorbed luminosity of the jet and disk ranges from 2 to 20 $\times$10$^{37}$~erg/s, with the disk reflection component contributing mainly to the hard 20--30 keV excess and the stationary 6.7~keV ionized Fe line complex.}
   {At low opening angles $\Theta$ we find that the jet expands sideways following an adiabatic expansion of a gas with temperature $T_o$. 
   Finally, the central source and lower parts of the jet could be hidden by an optically thick region of $\tau > 0.1$ and size $R\sim N_H/n_{e0}\sim1.5\times10^9$~cm$\sim$1700~$r_g$ for $M_{BH}=3~M_{\odot}$.}

    \keywords{X-rays: individual -- SS~433;
          X-rays: binaries   --
          microquasars --
          jet  -- emission processes}

   \maketitle

\section{Introduction} \label{sec:intro}

SS~433 is a Galactic eclipsing X-ray binary (XRB) system, member of the microquasar class \citep{Margon1984, Mirabel1998}. It is composed of an A-type supergiant star and either an accreting neutron star or a black hole \citep{Kubota2010, Robinson2017}, the nature of its compact object still being controversial, on a circular orbit with an orbital period of 13.1~days \citep{Fabrika2004}.  It seems to be located at a distance of $5.5 \pm 0.2$~kpc \citep{Lockman2007}, a value which is consistent with recent geometric parallax from \textit{Gaia} satellite ($4.6^{+1.9}_{-1}$ kpc at 1$\sigma$) \citep{Gaia2016}. 

Jets  in SS~433 are its more prominent feature. They are the most powerful ones known in the Galaxy with luminosities of $L_{jet}~\gtrsim~10^{39}$~erg~s$^{-1}$ \citep{Marshall2002}, and the first discovered for a compact Galactic source \citep{Abell1979, Fabian1979}. They are ejected at a mildly relativistic velocity of $v~\sim~0.26c$ \citep{Margon1989}. It is remarkable that baryons are present in these jets, SS~433 being together with 4U~1630--47 the only two Galactic  XRBs in which baryonic jets have been observed \citep{Kotani1994, DiazTrigo2013}. X-ray emission lines from ionized heavy elements have been detected \citep{Margon1989, Marshall2002}, associated to adiabatic expansion and radiative losses of hot and dense blobs of gas propagating outwards the compact source and following the jet  precessional motion.
Multiwavelength observations of the SS~433 outflow reveal a consistent scheme of symmetric jet flow, once Doppler boosting and projection effects are taken into account \citep{Roberts2010, Bell2011, Marti2018}, with adiabatic losses playing a major role in the jet emission, following a path accurately described by a kinematic model \citep{Hjellming1981, Margon1989}. Using
ALMA archival data, \cite{Marti2018} confirmed that the energy losses of radiating electrons in the jet are dominated by adiabatic expansion instead of synchrotron radiative losses.

Precession of the jet  in SS~433 has been extensively studied at different wavelengths for decades. Apart from its apparent shape, it was observed in both the Doppler-shifted X-ray with the EXOSAT satellite \citep{Watson1986} and optical \citep{Margon1979} emission lines, from which precessional parameters could be determined. The exhaustive monitoring of the source lead to the obtention of its power spectrum, thus allowing a time-series analysis which resulted on SS~433 being so far the only XRB with measured orbital, precessional and nutational period \citep{Eikenberry2001}. 

\cite{Medvedev2018} studied SS433 on the X-ray domain using data from \textit{Chandra} to describe the hard component of the spectra by including a hot extension of the jets, which is optically thick to low energy photons ($E<3$~keV) but progressively optically thinner to higher energy photons. This serves as a source of high dense absorption to the central source and lower parts of the jets, as well as an up-scattering component of soft photons emitted by the visible part of the jets. 

Although SS~433 has been extensively studied in the X-ray domain, data from \nustar\ satellite are not completely exploited yet. The \nustar\ observatory operates up to very hard X-ray energies ($3$ to $79$ keV) with spectral resolutions of 0.4~keV at 10~keV and 0.9~keV at 68~keV. The combination of emission line spectroscopy with the study of the hard X-ray continuum emission should thus provide a more detailed description of SS~433.

In this article we present a spectral analysis of a publicly-available dataset consisting of ten \nustar\ observations of SS~433, performed between October 2014 and July 2015. 
The paper is structured as follows: we present the observations and data reduction in \hyperref[sec:data]{Sect.~\ref{sec:data}}. In \hyperref[sec:results]{Sect.~\ref{sec:results}}, we show the results of our X-ray spectral analysis in the context of a kinematic model for SS~433 precessing jet. Finally, in \hyperref[sec:discussion]{Sect.~\ref{sec:discussion}} and \hyperref[sec:conclusions]{Sect.~\ref{sec:conclusions}} we draw and present our main conclusions derived from our results.

On a recent paper, \cite{Middleton2021} had also exploited the same \nustar\ dataset. In their work, they focused on the analysis of the time-resolved covariant spectrum, as well as the associated frequency- and energy-dependent time lags, which they used to constrain physical properties of the accretion regime, associated to different scenarios for SS~433. In our work, we analyze the complete dataset of 10 observations, without disregarding any of them, and we focus on the time-averaged, or stationary spectra, which we use to derive geometrical and physical properties of SS~433 using detailed jet and disk-reflection models, in the context of their mutual precessional motion.

\section{Data analysis} \label{sec:data}

\begin{table*}
    \renewcommand{\arraystretch}{1.5} 
    \centering
    \begin{tabular}{c c c c c c c c c}
        \hline
        Obs & Mode & MJD & Exp. [ks] & $\Psi_{\rm orb}$ & $\Psi_{\rm pre}$ & $\Psi_{\rm nut}$ & SAA parameters & Src/Bkg radii \\
        \hline
        02 & 01 & 56934.13 & 26.7 & 0.28 & 0.69 & 0.74 & Strict - Yes & 50" / 100" \\
        04 & 01 & 56960.35 & 25.3 & 0.28 & 0.85 & 0.91 & Strict - Yes & 50" / 100" \\
        06 & 01 & 56973.40 & 29.2 & 0.28 & 0.93 & 0.99 & Strict - Yes & 70" / 70" \\
        08 & 01 & 56986.44 & 27.8 & 0.28 & 0.02 & 0.06 & Strict - Yes & 70" / 70" \\
        10 & 06 & 56999.55 & 12.6 & 0.28 & 0.10 & 0.14 & Strict - Yes & 60" / 85" \\
        12 & 01 & 57077.93 & 21.4 & 0.27 & 0.58 & 0.61 & Strict - Yes & 70" / 70" \\
        14 & 01 & 57092.04 & 26.2 & 0.35 & 0.66 & 0.85 & Strict - Yes & 70" / 70" \\
        16 & 01 & 57104.74 & 29.5 & 0.32 & 0.74 & 0.87 & Strict - Yes & 70" / 70" \\
        18 & 01 & 57130.75 & 27.4 & 0.31 & 0.91 & 0.01 & Strict - Yes & 70" / 70" \\
        20 & 01 & 57208.00 & 26.6 & 0.21 & 0.38 & 0.30 & Strict - Yes & 70" / 70" \\
        \hline
    \end{tabular}
    \caption{\nustar\ observations of SS~433. Obs. column contains shortened names for ObsIDs 300020410\#\#. Modes 01 and 06 correspond to Science, and Spacecraft modes, respectively. Southern Atlantic Anomaly (SAA) parameters, and Source and Background extraction radii are also included. Orbital ($\Psi_{\rm orb}$), precessional ($\Psi_{\rm pre}$), and nutational phases ($\Psi_{\rm nut}$) were calculated based on ephemeris of \cite{Eikenberry2001}.}
    \label{tab:obstable}
\end{table*}

\nustar\ observed SS~433 for 10 times between modified Julian dates (MJD) 56934 and 57207 with typical exposures of 20--30~ks in the 0.2--0.3 orbital phase range, spanning over roughly one and a half precessional periods of the source. Details of the observational dataset are given in \hyperref[tab:obstable]{Table~\ref{tab:obstable}}. Observations ID span from 30002041002 to 30002041020. From now on, we shorten their Obs names to the last two digits for simplicity. Due to the triggered read-out mechanism of \nustar, the spectra derived for a source as bright as SS~433 have a great signal to noise ratio and are safe from pile-up.

We processed the data obtained with the two Focal Plane Modules \citep[FPMA and FPMB;][]{Harrison2013} using the \nustar\ Data Analysis Software ({\sc NuSTARDAS}) available inside {\sc HEASOFT}~v6.28 package. The observation files were reduced with the {\sc nupipeline} tool using the CALDB~v.20200429. 
We generated source and background spectra, as well as the ancillary and response matrices for each observation using the {\sc nuproducts} script. We extracted photons in circular regions of 50 to 70~arcsec centered at the centroid of the source and of 70 to 100~arcsec for the background, using the same chip, in regions that were not contaminated by the source. 
The X-ray spectral analysis was performed using \xspec~\citep{Arnaud1996} considering the 3--70 keV energy range, as we did not detect significant emission from the source over the background level at higher energies.

In order to filter the Southern Atlantic Anomaly (SAA) passages we applied different criteria depending on the individual observation reports\footnote{\href{http://www.srl.caltech.edu/NuSTAR\_Public/NuSTAROperationSite/\newline SAA\_Filtering/SAA\_Filter.php}{SAA reports}}.
We also performed the standard analysis {\sc SAAMODE=none} and {\sc TENTACLE=no} and checked the dependence of our results on this filtering. For each observation, we found that the spectral parameters were consistent within the errors. We performed a similar check by considering two different spectral backgrounds, and we obtained consistent results.

For Obs10, the total exposure of Science Mode 01 was about 3~ks. We thus reduced the Spacecraft Mode~06 data by means of the standard splitter task {\sc nusplitsc}. Using the Camera Head Unit {\sc CHU12} combination in {\sc STRICT} mode, we obtained an enhanced exposure of 12.59~ks, and we used this dataset for the spectral analysis.

In \hyperref[tab:obstable]{Table~\ref{tab:obstable}} we show the ten observations and their characteristics including the operating mode, MJD date, final GTI exposure, precessional, nutational and orbital phases as well as the SAA parameters used for GTI filtering and the extraction radii for the spectral analysis. Phases were calculated based on the ephemeris of \citet{Eikenberry2001} and include their corresponding intervals according to their exposure time fraction.

\section{Results} \label{sec:results}

\subsection{Model setup}
\label{sec:model}

In order to investigate the spectral X-ray variability of SS~433 along the ten \nustar\ observations, we propose the same spectral model for the whole set of averaged spectra, with similar Galactic absorption, jet  and accretion disk components. 
In particular, we consider a double neutral Galactic absorption model \tba\ with abundances of \citet{angr} and \citet{bcmc} cross sections. We fixed the Galactic absorption parameter through all of our spectral analysis to a value of $N_H=0.67\times 10^{22}$~cm$^{-2}$ \citep{Marshall2002, Namiki2003} while leaving local absorption free. In \xspec\ this double absorption component reads \tba \texttt{*}\tba.

To account for cross-calibration uncertainties between both \nustar\ instruments FPMA and FPMB, we include a constant factor between each spectrum (\cons\ model in \xspec). 
We checked that this constant remains in the 3\% level for all the epochs,  which is inside the expected 0--5\% range \citep{Madsen2015}. The spectra were grouped to a minimum of 30 counts per bin to properly use $\chi^2$ statistics. Throughout all this paper we quote parameter uncertainties to 90\% confidence level, computed using \xspec\ {\sc chain} task with Goodman-Weare algorithm and 360 walkers (20 times the number of free parameters).

To check for convergence of MCMC chains, we visually inspected the chains of each parameter and determined the most appropriate number of burn-in steps in order to obtain uncorrelated series for the parameters of interest.
We corroborated this method by computing the integrated autocorrelation time associated with each series, and verified that it remained as close as unity as possible (see \href{https://emcee.readthedocs.io/en/stable/tutorials/autocorr/}{documentation} on the {\sc python-emcee} package for more details).
We found that a total length of $1.2\times10^7$ with a burn-in phase of $6\times10^6$ was sufficient for all ten observations to reach convergence.

On \hyperref[fig:trace]{Figure~\ref{fig:trace}} we show an example of a parameter chain series with the computed integrated autocorrelation time $\tau$. 

\vspace{0.5cm}

To represent the X-ray emission from the jet in SS433, we considered a spectral model developed by \cite{Khabibullin2016}. We adopted the SS433 flavour which has the jet  opening angle and bulk velocity fixed at $\Theta=0.024$~rad and $\beta=0.2615$ respectively. This table model depends on the jet kinetic luminosity $L_k$, the jet base temperature $T_o$ and the electron transverse opacity $\tau_{e0}$ at the base of the jet. The model also includes iron and  nickel  abundances. By considering a distance of $d_s=5.5$~kpc \citep{Blundell2004}, the model normalization can be expressed as $ N = L_{38} \tau_{e} / d_{10}^2 $, where $L_{38}=L_k/10^{38}$~erg~s$^{-1}$ and $d_{10}=d_s/10$~kpc. 

To account for the western jet contribution, we included a second additive table model multiplied by an attenuation factor $C_{\rm west}^{\rm jet}$ (\cons\ in \xspec). The  western jet parameters were linked to that of the eastern jet. Both table models \textbf{were} loaded in the \xspec\ environment by means of the \textrm{atable} command. To account for the precessional motion of the jet  (Doppler shifting and boosting) we included a convolution model (\zsh\ in \xspec) to each table model. A Gaussian smoothing component \textrm{gsmooth} (with index $\alpha=1$) was also added to take into account broadening of the emission lines caused by the gas expansion of the ballistic jet . Therefore, both jet  X-ray spectra are modelled by \zsh*\bjet+\cons*\zsh*\bjet, in \xspec\ language.

To account for the accretion disk emission, we included a linear combination of direct thermal emission from a blackbody spectrum (\dbb), which contributes significantly at energies below 5 keV, and pure-reflected neutral (\xcp; \citet{Garcia2013}) and relativistic-emission spectra (\rlp; \citet{Dauser2014}). In both latter components, we chose to use the coronal flavours (\textrm{Cp}). In the relativistic case, we chose the lamp post geometry (\textrm{lp}).
These reflection components contribute both to the ionised iron clomplex at $\sim$6.7~keV, and to the hard excess at 20--30 keV through the Compton hump.

In summary, the complete disk emission spectrum is modelled by \dbb+\xcp+\rlp. The free parameters are: the temperature $kT_{\rm dbb}$ and the normalization of the blackbody component $N_{\rm dbb}$; the incident photon spectrum index $\Gamma$, the ionization degree $\xi$, the inclination angle $\phi$ and normalization of the \xcp\ component; the source height $h$ above the disk of the \rlp\ component. The reflection fraction was set to $-1$ in order to obtain only the reflected spectrum.
The iron abundance and coronal temperature of the \xcp\ component were tied to their respective analogues of the \bjet\ components.
All identical parameters of both reflection components where tied together. The remaining parameters were left frozen to their default values.

The resulting best-fitting parameters of the entire model are shown on \hyperref[tab:modparams]{Table~\ref{tab:modparams}}. On \hyperref[fig:squema]{Figure~\ref{fig:squema}} we show a simplified picture of the SS433 jet-disk system, indicating each model contribution to the total X-ray spectra.

\begin{figure*}
	\centering
	\includegraphics[width=0.45\textwidth]{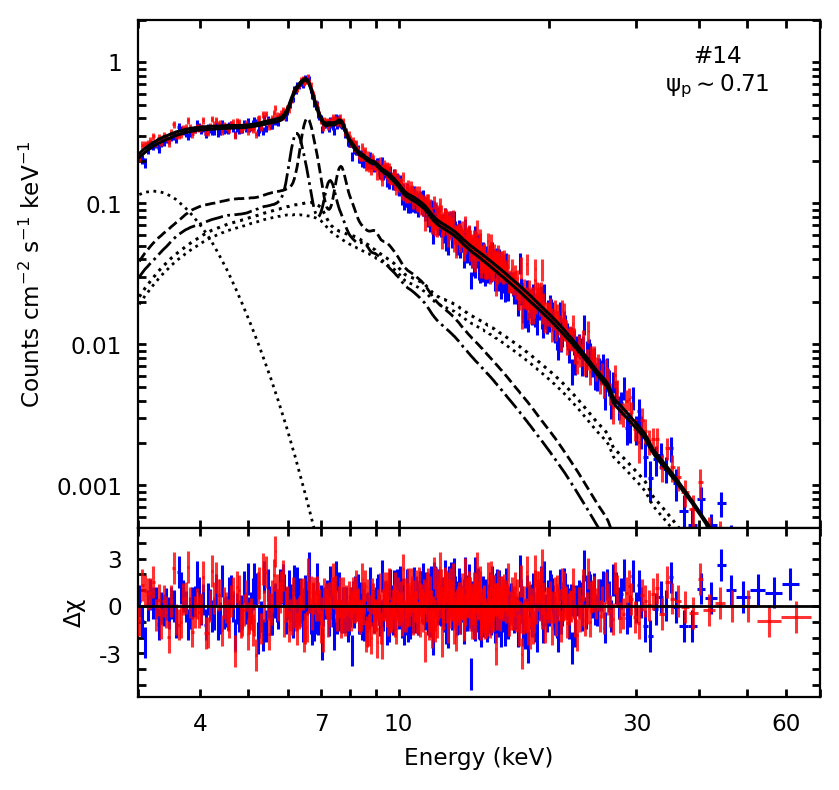}
	\includegraphics[width=0.45\textwidth]{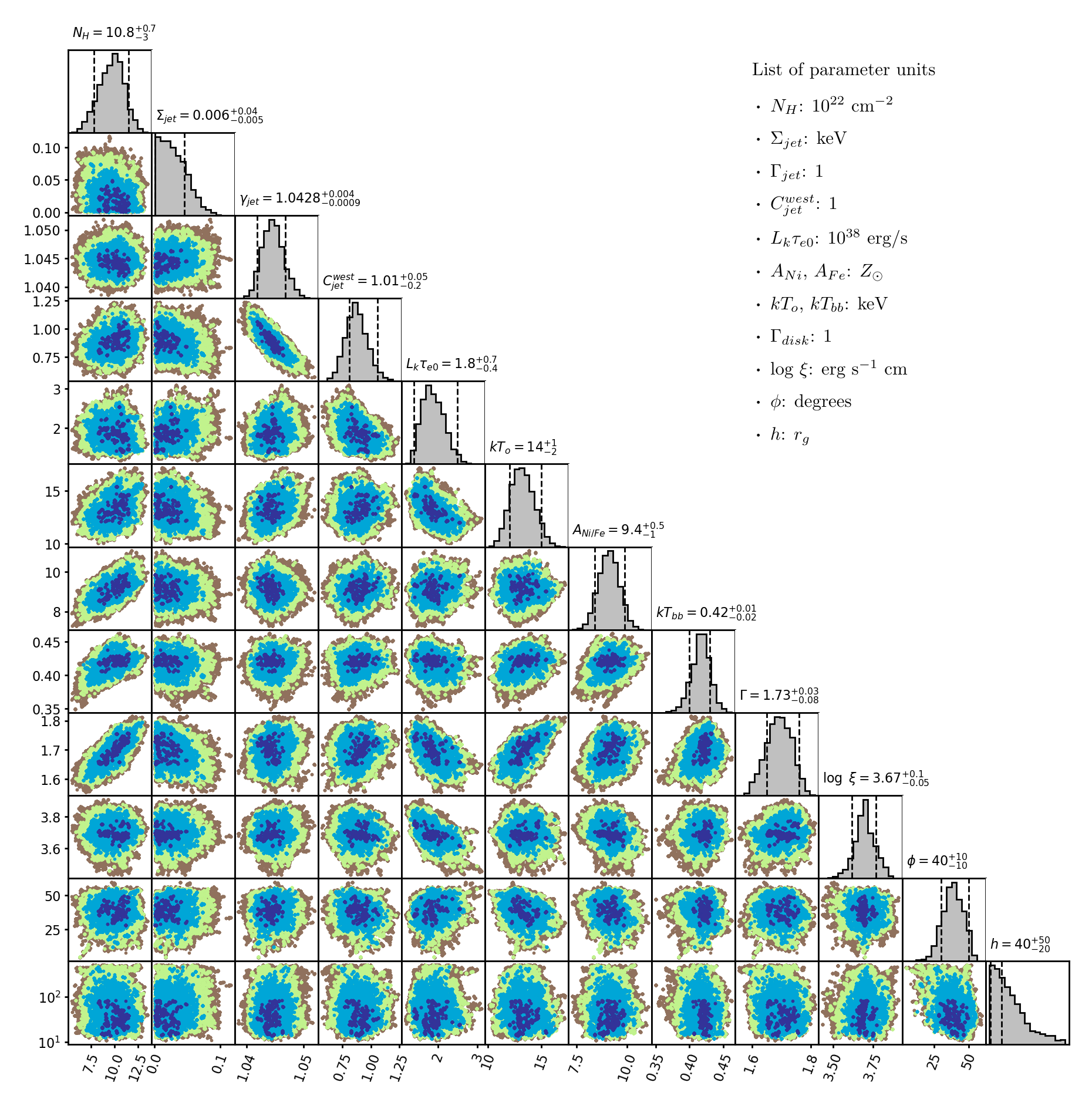}
	\includegraphics[width=0.45\textwidth]{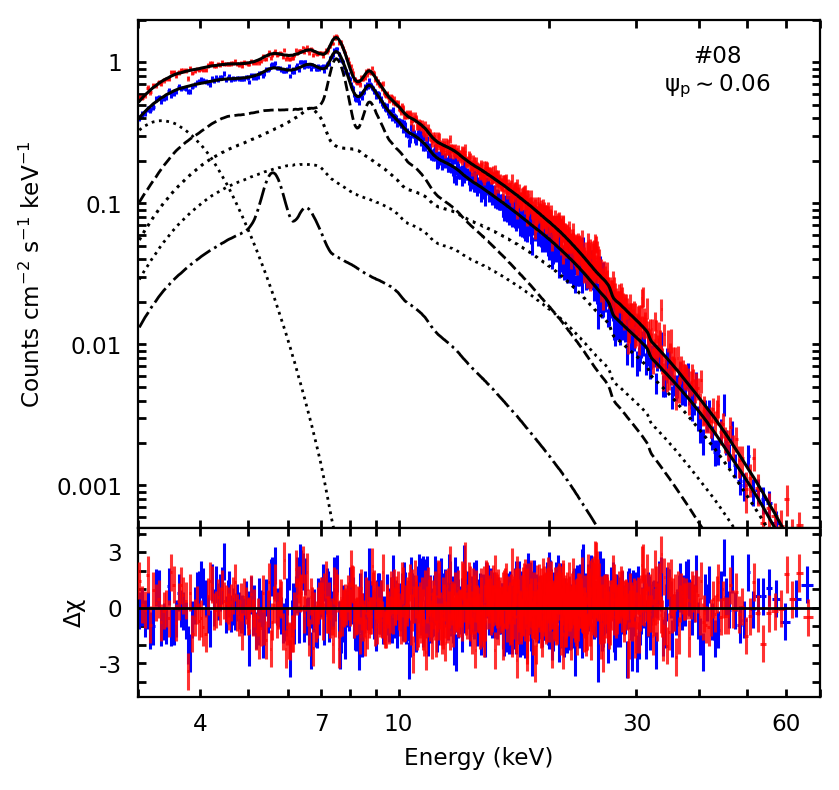}
	\includegraphics[width=0.45\textwidth]{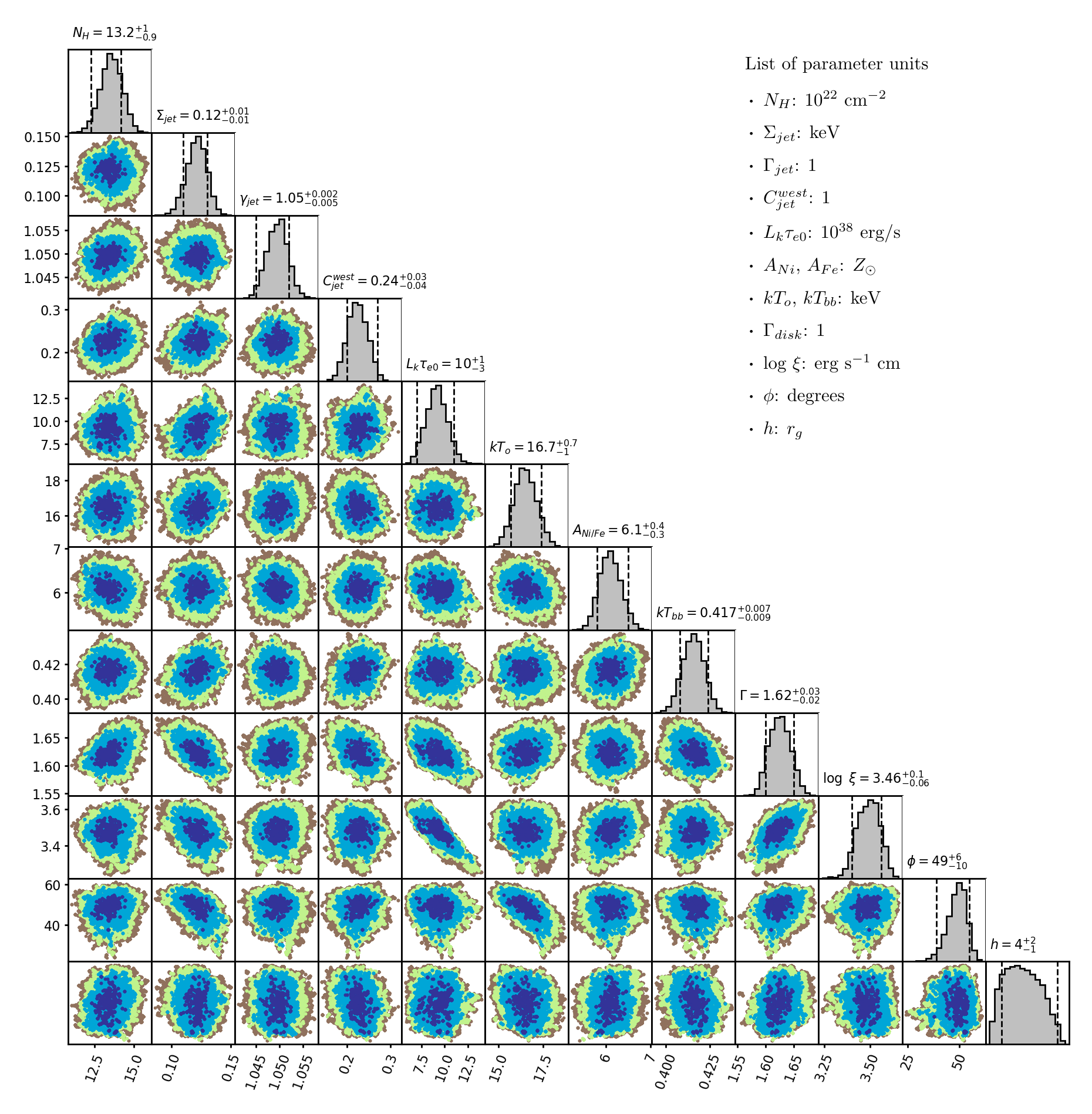}
    \caption{Left column: Sample of \nustar\ FPMA/B averaged spectra fitted with a combination of \bjet, \dbb, \xcp\ and \rlp. Dashed line corresponds to the eastern jet model component. Dot-dashed line corresponds to the western jet model component. Dotted lines corresponds to the different disk model components. 
    Right column: one and two dimensional distribution of some continuum model parameters derived from the MCMC simulations. Outwards colour gradient indicate confidence levels: 90\%, 99\%, 99.9\% and 99.99\%}. See \hyperref[tab:modparams]{Table~\ref{tab:modparams}} for details on parameters units.
    \label{fig:specs}
\end{figure*}

\subsection{Broadband description}

On the left panels of \hyperref[fig:specs1]{Figure~\ref{fig:specs}} we show observations \#14 ($\psi_{\rm pre}\sim0.71$) and \#08 ($\psi_{\rm pre}\sim0.06$) spectra and their best fits along with their residuals. These two examples show two very different instances of the precessional motion. Observation \#14 has both jets at similar Doppler shifts and thus showing overlapping emission lines (unresolvable by \nustar). Observation \#08 has the eastern and western jet at opposing Doppler shifts, and thus showing iron and  nickel  emission lines perfectly resolvable by \nustar\ (dashed and dot-dashed lines).
We also clearly see the different disk component contribution (dotted lines) at very soft energies (E$<5$ keV; \dbb), the Fe~K$\alpha$ line ($\sim$6.4~keV; \xcp) and the harder (E>20~keV) reflected component (both \xcp\ and \rlp).

The soft energy range of the \nustar\ spectra (E$<10$~keV) is dominated by the contribution of one or both jets and the thermal disk component. 
On highly blue-shifted phases ($\psi_{\rm pre}<$0.2 and $\psi_{\rm pre}$>0.8), the western jet contribution to the total flux seems to be $\sim$0.1--0.3 times that of the eastern jet, as modelled by the attenuation factor. During the in between phases, when the merging of emission lines starts to occurs, the western jet contributes significantly more, with factors ranging from 0.6 to 1. 

\begin{figure*}
	\centering
	\includegraphics[width=0.48\textwidth]{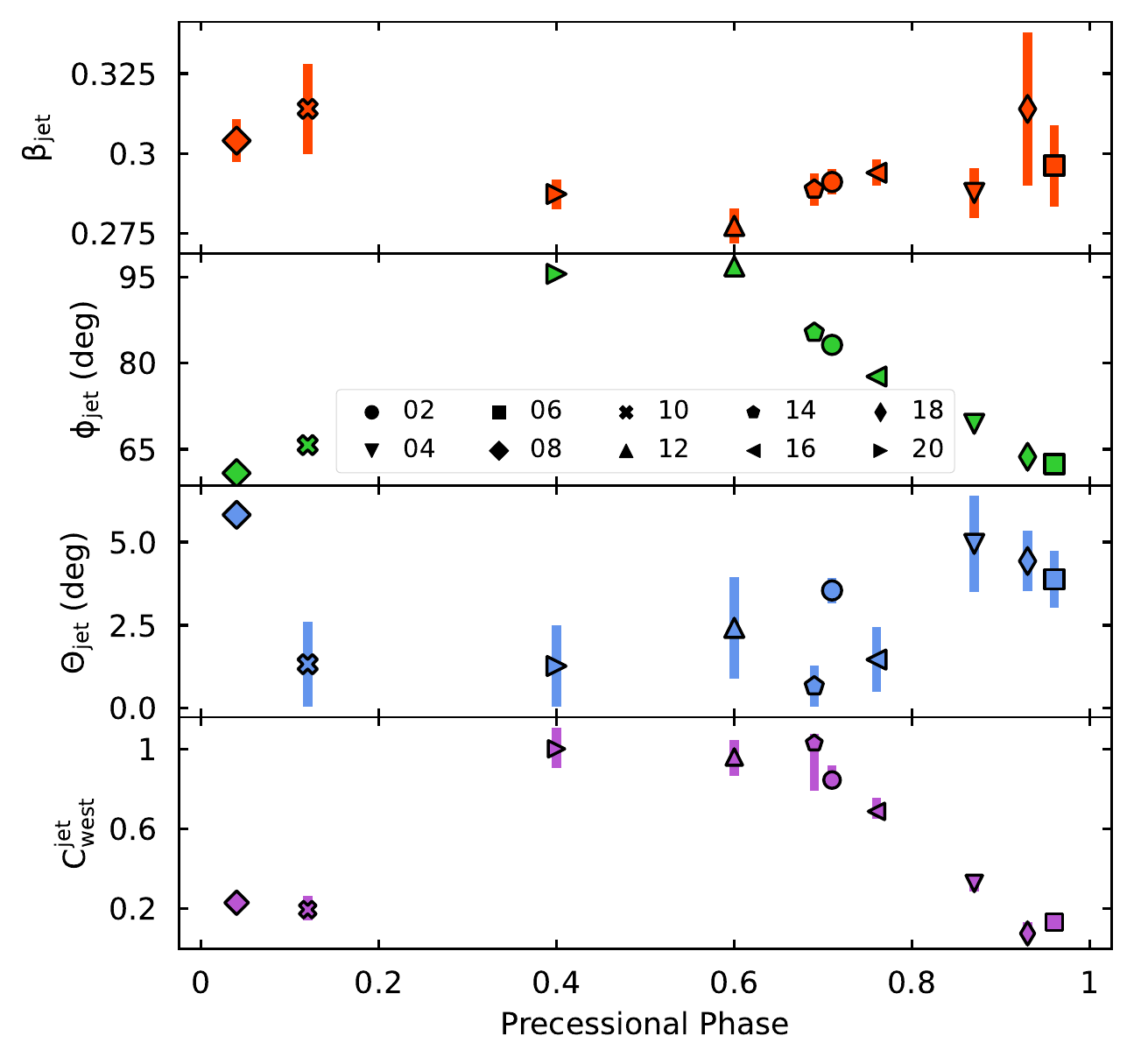}
	\includegraphics[width=0.48\textwidth]{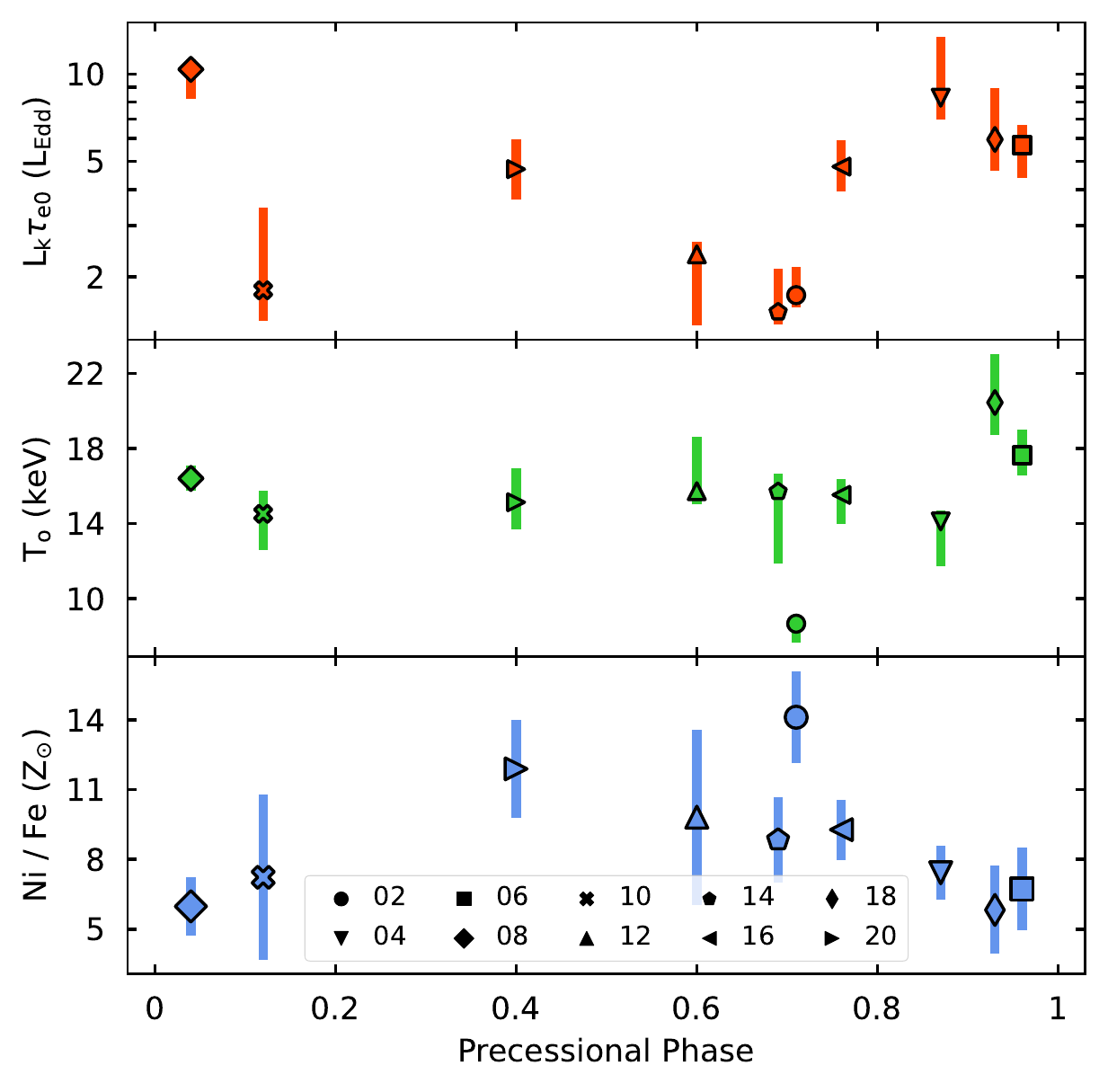}
    \caption{Precessional evolution of \bjet\ and system parameters derived from the 3--70~keV fit to \nustar\ spectra. 
    Left panel, top to bottom: bulk velocity factor $\beta=v/c$, jet inclination angle $\phi$, jet opening angle $\Theta$, western jet attenuation factor.
    Right panel, top to bottom: jet kinetic luminosity weighted by opacity $L_{k}\times\tau_{e0}$ ($10^{38}$~erg~s$^{-1}$), jet base temperature $T_{o}$ and ratio of  nickel  to iron abundances.}
    \label{fig:jetpars}
\end{figure*}

\begin{figure*}
	\centering
	\includegraphics[width=0.48\textwidth]{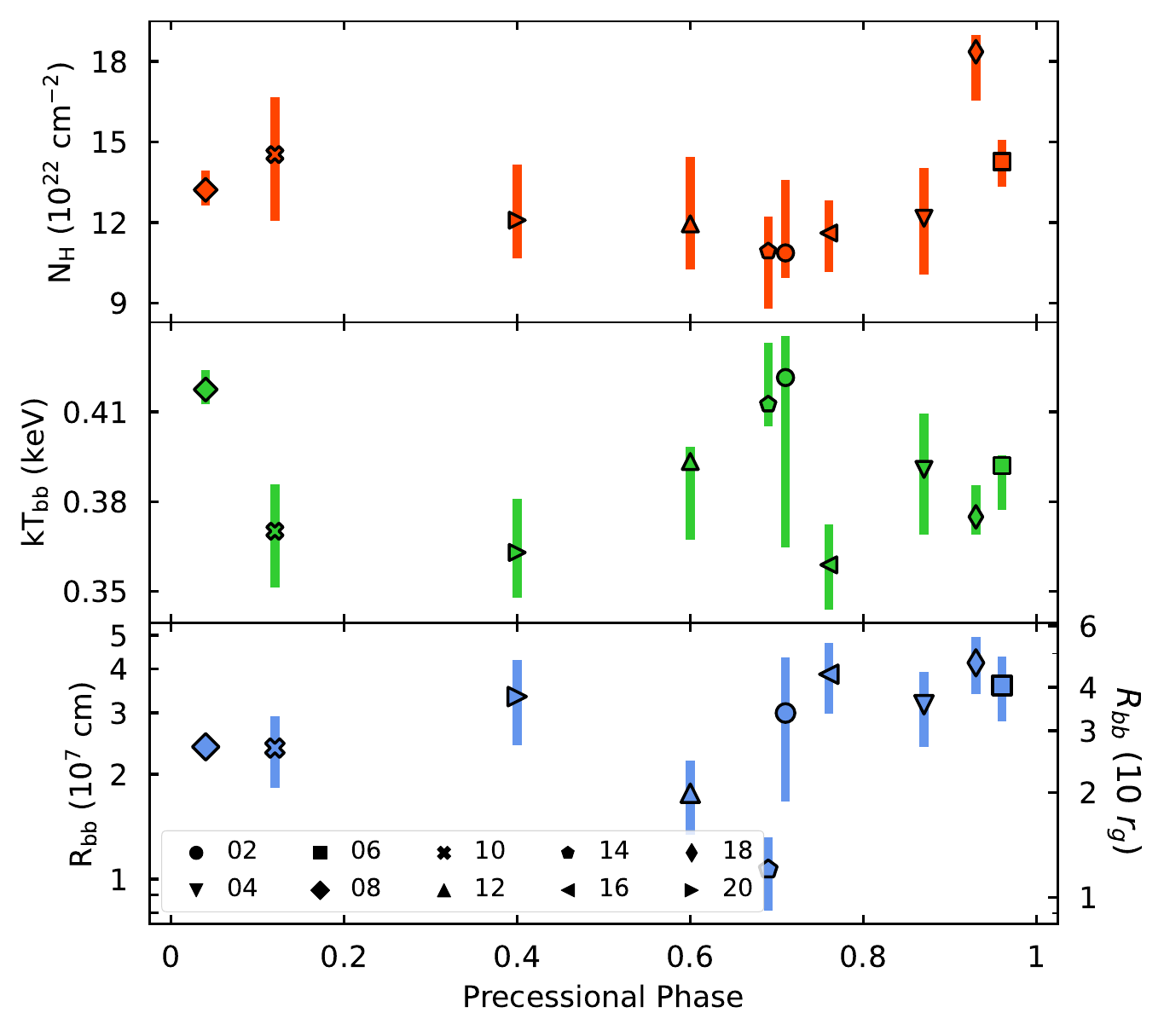}
	\includegraphics[width=0.48\textwidth]{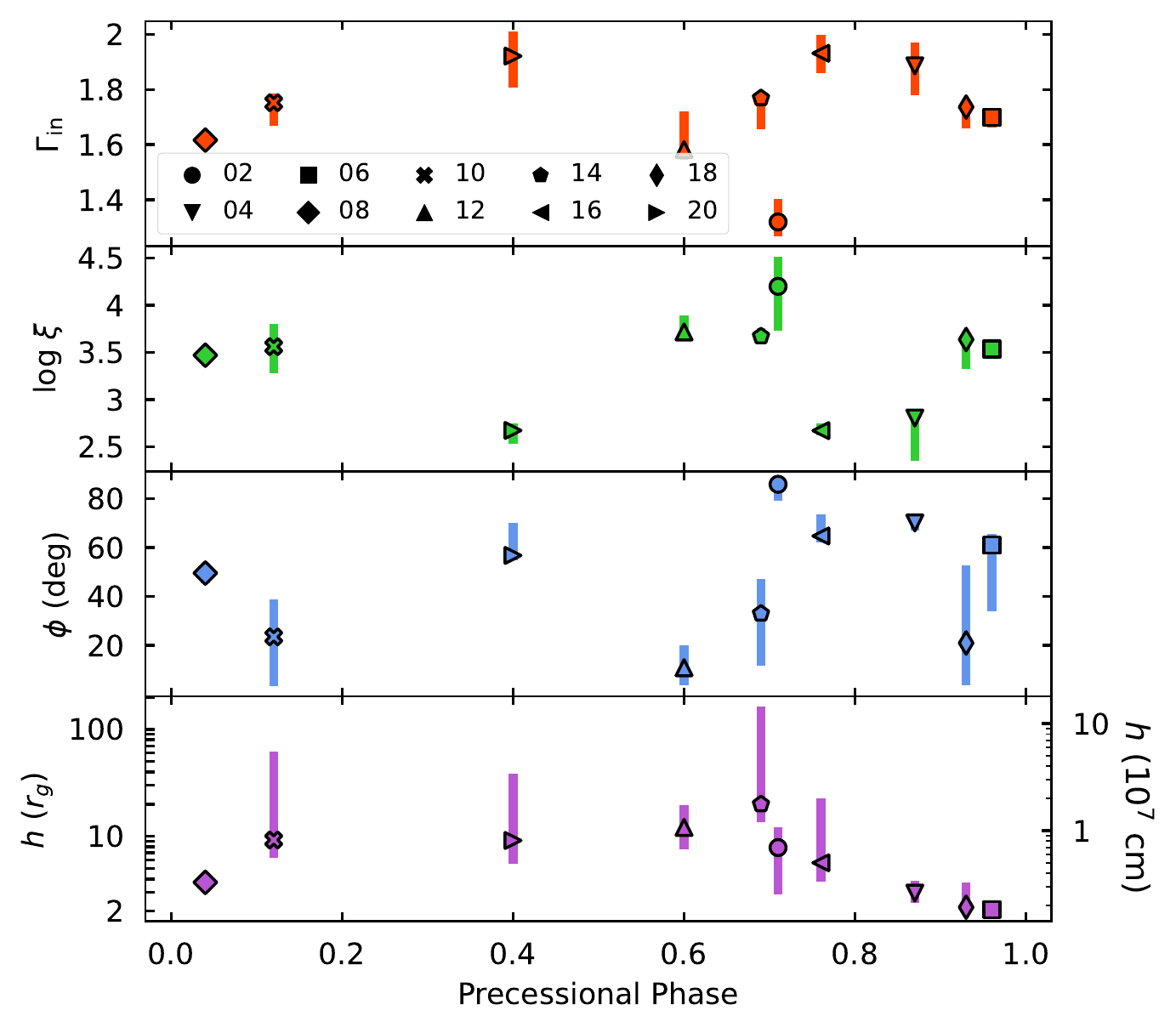}
    \caption{Precessional evolution of different disk components parameters.
    Left panel, top to bottom: local absorption column, \dbb inner radius temperature and inner radius.
    Right panel, top to bottom: incident spectral index, ionization degree, inclination angle and source height.
    }
    \label{fig:diskpars}
\end{figure*}

The absorption column density does not seem to vary significantly among the different precessional phases. It stays somewhat high and constant at an average value of $12 \times 10^{22}$~cm$^{-2}$. We must note that \nustar\ lower energy detection limit of 3 keV does not allow to better constrain this parameter. Furthermore, the black body component also dominates at very low energies, so the absorption column and black body parameters (temperature and normalization) are tightly correlated (see the left panel of \hyperref[fig:jetpars]{Figure~\ref{fig:diskpars}}).

Using the thermal \dbb\ component we get an inner temperature than ranges from approximately 0.36 to 0.42 keV. This model normalization $N_{\rm \dbb}$ can be used to estimate the inner disk radius $N_{\rm \dbb} = (f_{c}~R_{\rm \dbb}/d_{10})^2~\cos\phi$, where $f_c = 1.7$ is the color-temperature correction factor \citep{Kubota1998}, 
and $\phi$ is the angle between the normal to the disk and the line of sight. 
As shown on the bottom left panel of \hyperref[fig:jetpars]{Figure~\ref{fig:diskpars}}, we see that this parameter remains very well constrained between $1-5 \times 10^7$~cm (10--60 $r_g$ for $M_{BH}=3~M_\odot$).

As already mentioned on the previous section, the \bjet\ model normalization can be expressed in terms of $L_k$ and $\tau_{e0}$ by fixing a distance to the source. 
We constrain the value of $L_{k}*\tau_{e0}$ and use it as a measure of the jet kinetic power \citep{Khabibullin2016} transverse to the outflow axis. 
We report the best \bjet\ parameters on the right panel of \hyperref[fig:jetpars]{Figure~\ref{fig:jetpars}.}.
The jet kinetic luminosity ranges between $1-10$ times the Eddington luminosity ($\sim$10$^{38}$~erg~s$^{-1}$), with higher values at extreme precessional phases. The temperature at the base of the jet (where it becomes visible in X-rays) ranges from 12 to 18~keV (within errors), averaging $\sim$15 keV. The base electron optical depth ranges between $\sim$0.1 and its maximum accesible value of 0.5.

We notice the nickel  overabundance with respect to iron already reported on \cite{Medvedev2018}. The  nickel  to iron abundance ratio varies between 5--15, being highest on the intermediate phases. Although an apparent precessional motion of this ratio can be seen (bottom right panel of \hyperref[fig:jetpars]{Figure~\ref{fig:jetpars}}), it can be thought to be constant at $\sim$9 within 1$\sigma$.

The disk continuum parameters show a more intricate behaviour. The incident power-law index ranges between 1.6 to 2, taking the lowest value of $\sim$1.4 at observation \#02. It becomes harder towards intermediate phases, pointing to a weaker jet dominated state.
It is interesting to note that the spectral index of the \texttt{pexmon} component used by \cite{Middleton2021} to fit the average spectra, which was tied across all 8 observations, is considerably harder than the one found in the fits to the covariance spectrum ($\sim$1.4 and $\sim$2.2 respectively).
Our results lie between these two boundaries, which shows that our precessional analysis is compatible with the fits to the time-resolved covariance spectra.

Both ionization degree $\xi$ and inclination angle $\phi$ do not seem to follow any particular precessional behaviour, but instead seem somehow anti-correlated. At lower inclination angles, the ionization degree increases. This may indicate that at lower (higher) inclinations we see more (less) of the inner and hotter regions of the accretion disk, and thus reflected on a higher (lower) ionization degree of the reflecting material.

Lastly, the illuminating source height $h$ ranges (within errors) between $0.2-9 \times 10^{7}$~cm (3--100~$r_g$ for $M_{BH}=3M_{\odot}$), taking lower values (with lower relative errors) towards extreme phases. 
This effect might be related to the fact that the more edge-on the accretion disk is seen, the weaker is the contribution to total flux from reflection, and thus, the most important is the contribution of direct emission.
This makes the height parameter more difficult to constrain.

On the right column of \hyperref[fig:specs]{Figure~\ref{fig:specs}} we present the triangle plot of observations \#14 and \#08, where the diagonal subplots represent the one dimensional distribution (histogram) of each parameter derived from the MCMC chains. The remaining subplots contain the two dimensional distribution of values of the $i$-th column parameter with the $j$-th row parameter. Colours indicate different confidence levels: 90\% (red), 99\% (green) and 99.9\% (blue).
For better display, we show only a subset of parameters. The top label above each parameter histogram indicates the relevant parameter name and its best fit value with the 90\% confidence level error range. We also included a table of units for clarity.

\subsection{Flux evolution and hardness}

\begin{figure*}
	\centering
 	\includegraphics[width=0.48\textwidth]{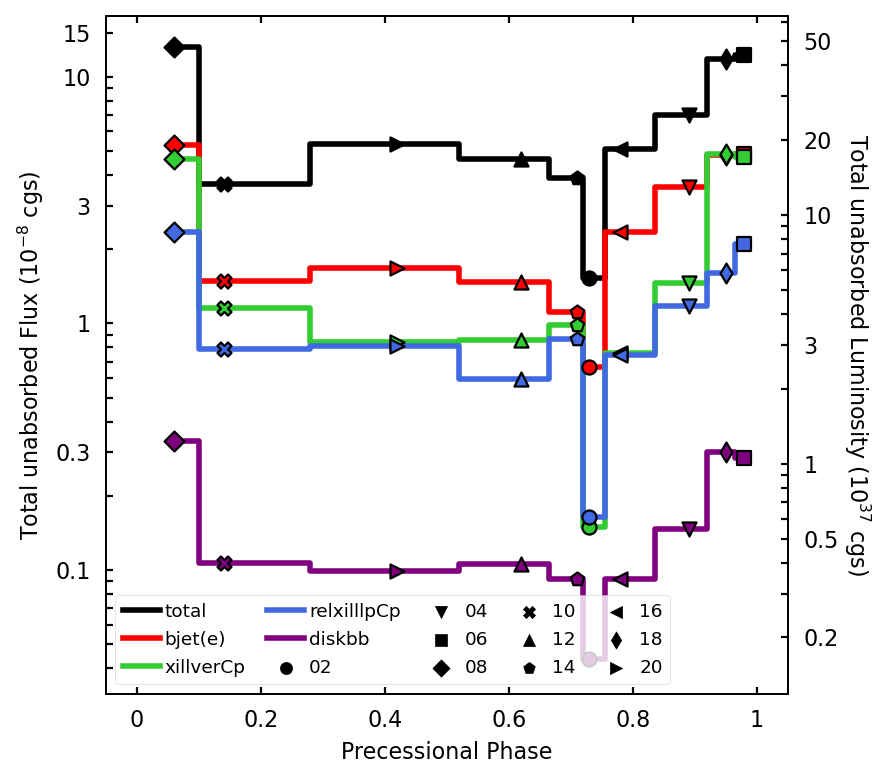}
	\includegraphics[width=0.48\textwidth]{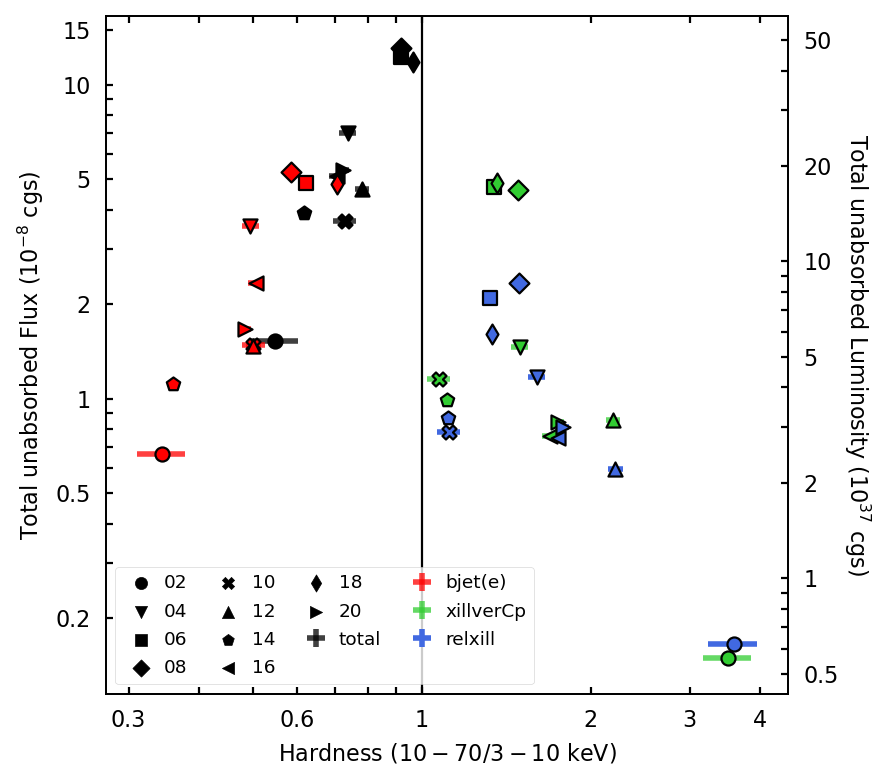}
    \caption{Spectral behaviour of the jet and disk model components. 
    Left panel: precessional evolution of the total flux (3--70 keV) of each model component. 
    Right panel: hardness-intensity diagram (soft: 3--10 keV, hard: 10--70 keV). 
    The \dbb\ (purple) component was only included on the left panel.
    Error bars appear smaller than the marker size on almost all data points.}
    \label{fig:fluxes}
\end{figure*}

To further investigate the spectral contributions of the jet  and the accretion disk along the precessional motion, we calculated each model unabsorbed flux (using \texttt{cflux} convolution model in \xspec) in 2 different bands: soft: 3--10~keV and hard: 10--70~keV. 

On the left panel of \hyperref[fig:fluxes]{Figure~\ref{fig:fluxes}} we show the precessional evolution of both defined bands for each model component. We only show the "eastern" \bjet\ component, as the remaining "western" would be the same multiplied by the attenuation factor. We also show the total flux for reference. 

The eastern jet component dominates and contributes from 30\% up to 65\% of the total observed flux depending on the precessional phase. 
The thermal disk component contributes $\sim$2\% almost independently of phase to the total flux.

On intermediate phases, where the total flux is reduced by a third, the contribution of both jet dominates, while the contribution of the disk comes almost equally from \xcp\ and \rlp\ components. On this phase (0.15--0.85) the disk components contribute up to 30\% of the total flux.

On the extreme phases, the total flux is distributed as follows: 65\% from the disk (considering the 3 components) and 45\% from both jet  (mainly the eastern jet). Moreover, the neutral reflection component has almost the same flux as the eastern jet component. This could be attributed in part to the beaming effect produced by the particular orientation of the system on these phases.

Lastly, we note a similar precessional behaviour between the jet and disk measured fluxes, the attenuation factor and the illuminating source height. 
The more edge-on the accretion disk is, the less contribution to total flux from reflection there is, and thus, the more important is the contribution of the direct emission.
On these precessional phases, the western jet emission becomes significant also.

When looking at the spectral distribution within energy bands on the right panel of \hyperref[fig:fluxes]{Figure~\ref{fig:fluxes}}, we see a clear difference between systems. 
By defining the hardness as the ratio between the measured fluxes of 10--70 keV (hard) to 3--10 keV (soft), we see that the jet component is purely soft X-ray dominated, while the disk components (without the thermal \dbb) is purely hard X-ray dominated. We also note that as total flux increases, every component tends to a hardness ratio of 1. Inversely, as total flux decreases, the jet becomes softer and the disk harder. 

As a final remark, we note that the total \bjet\ unabsorbed luminosity on the 3--70 keV band (assuming a distance of 5.5~kpc), ranges from $0.2-2 \times 10^{38}$~erg~s$^{-1}$. These values are approximately 2 to 10 times that of the measured kinetic luminosities (see \hyperref[fig:jetpars]{Figure~\ref{fig:jetpars}}). On the intermediate phases, the ratio between these two kind of luminosities is the greatest.

\section{Discussion}
\label{sec:discussion}

\subsection{Precessing lines and kinematic model}

\begin{figure*}
	\centering
	\includegraphics[width=0.475\textwidth]{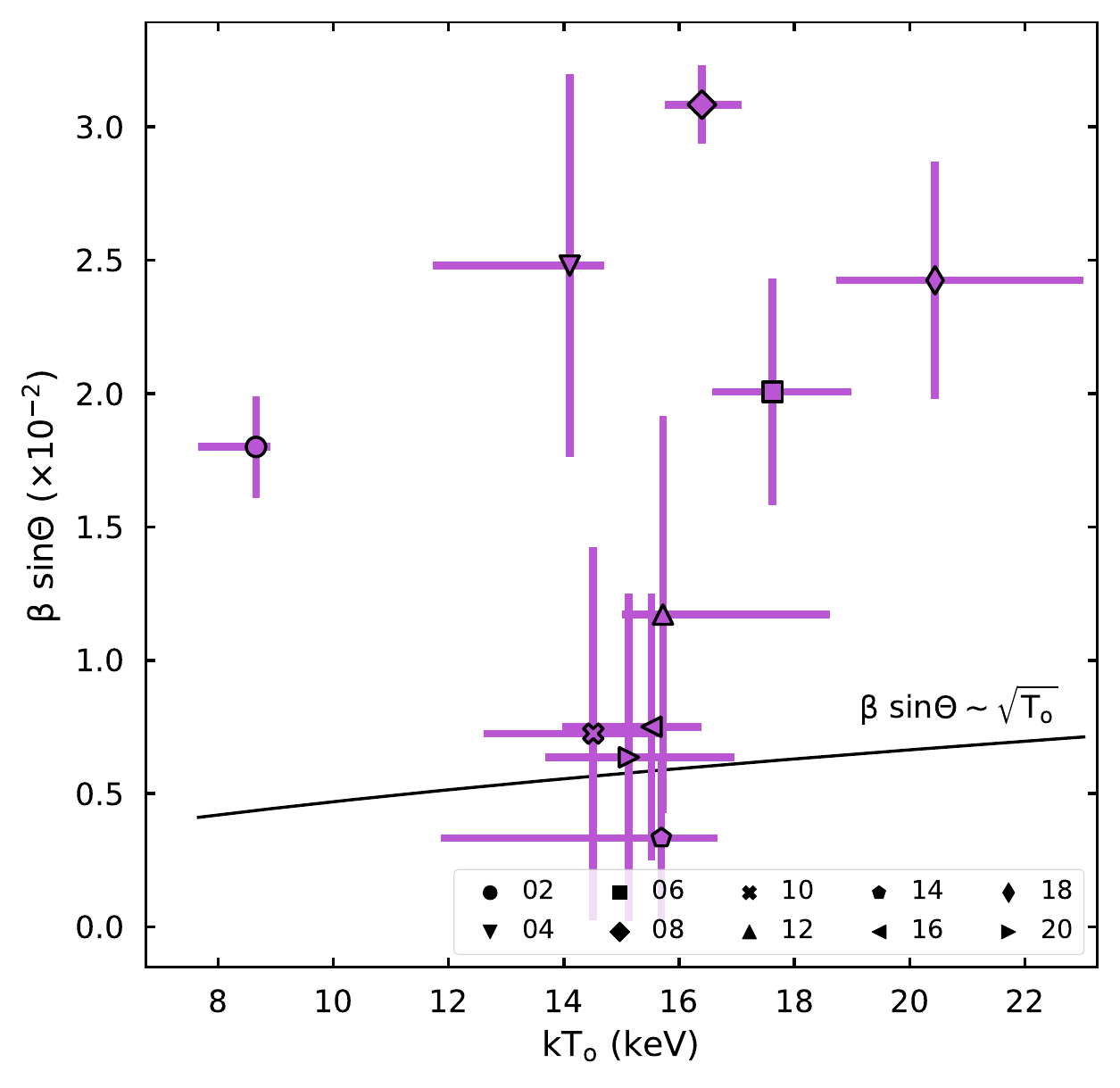}
 	\includegraphics[width=0.475\textwidth]{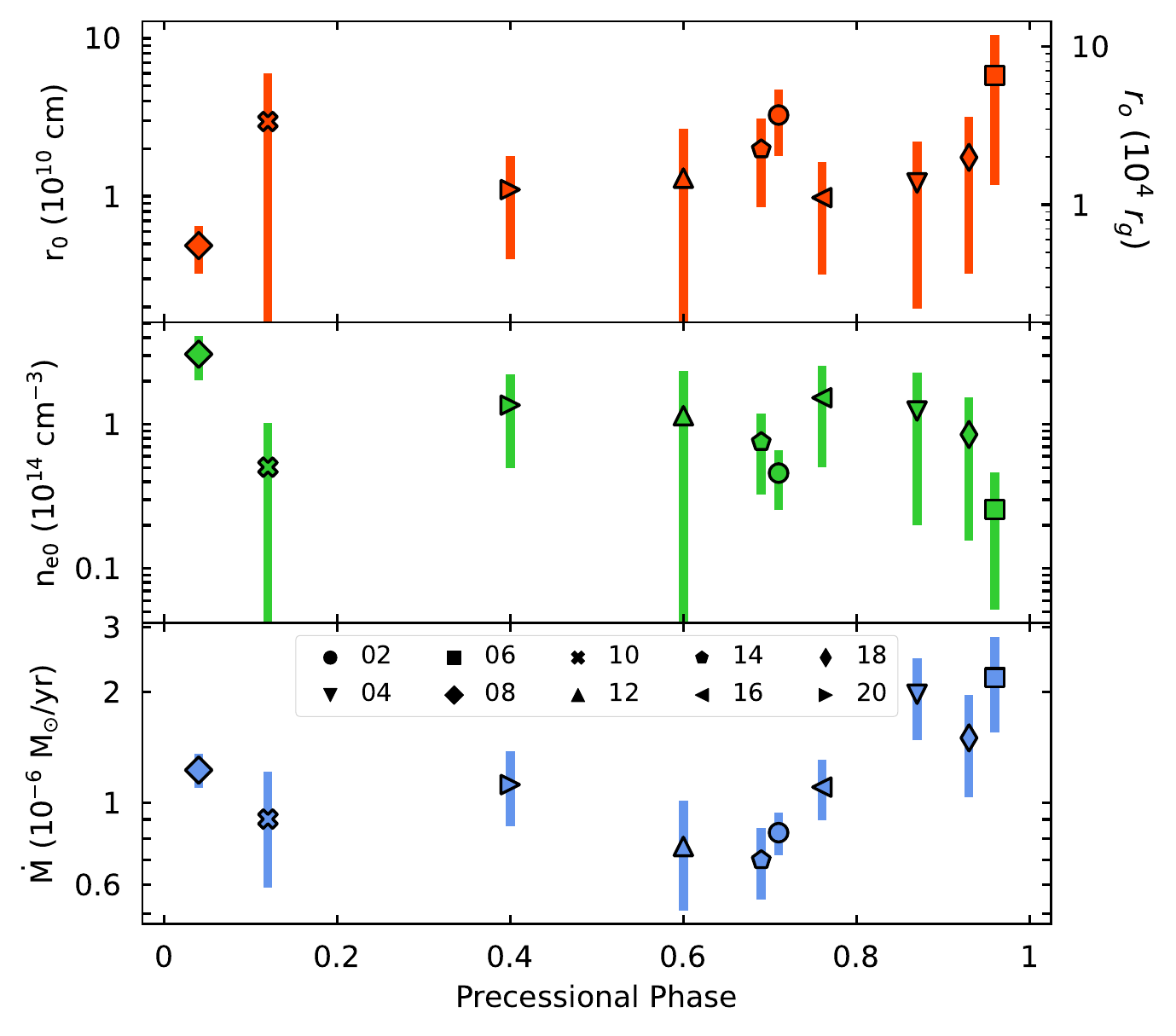}
    \caption{Left: Relationship between the temperature at the base of the jet and the transversal velocity of matter throught the jet. We plot the expected velocity caused by adiabatic expansion of the jet.
    Right: Precessional evolution of different jet parameters (see text for equations). From top to bottom: jet base height from cone apex, electron density at base height and mass flow rate through the jet.}
    \label{fig:betatemp}
\end{figure*}

From the obtained values of both jet redshifts $Z$ and the broadening factor of the lines $\Sigma$, we use the kinematic model equations in order to compute the bulk velocity of matter across the jet, $\beta_{\rm jet}=v/c$, the angle sustained by the eastern jet with respect to the observer $\phi$, and the half opening angle $\Theta$ of the jet.

Let us define $z_{e}$ and $z_{w}$ as the respective eastern and western jet redshifts. Then, by assuming perfect alignment between jets and equal velocities, we have:
\begin{equation}
    \beta=\sqrt{1-\frac{1}{(1+z_{o})^2}}~{\rm with}~z_{o}=\frac{z_{e}+z_{w}}{2}.
\end{equation}

By application of kinematic equations (see \citealp{Cherepashchuk2018} for full set of equations) we can express the angle between the jet axes and the line of sight in terms of both jet redshifts:
\begin{equation}
    \cos~\phi = \frac{z_w-z_e}{2\gamma\beta}~{\rm with}~\gamma=1+z_o~\rm{the~Lorentz~factor.}
\end{equation}

Lastly, by considering the line profiles to be Gaussians with dispersion $\Sigma$~($E_{o}$) (with $E_{o}$ the line centroid at rest) we can estimate the jet half opening angle \citep{Marshall2002}:
\begin{equation}
    \Theta = \sqrt{\frac{2~\log~2}{3}}~\frac{2}{\gamma~\beta~\sin\phi}~\frac{\Sigma(E_o)}{E_o}.
\end{equation}

The application of these three equations can be seen on the left panel of \hyperref[fig:jetpars]{Figure~\ref{fig:jetpars}}. 

Overall, we get an average bulk velocity factor $\beta\sim0.29$, and with values (and errors) that increase towards extreme phases. This effect comes from the fact that at higher redshift, the western jet redshift becomes harder to constrain, as its flux becomes significantly lower than the eastern jet, and competes with the thermal \dbb and the reflection components.
For comparison, the reference value obtained from decades of optical data is of $\sim0.26$ \citep{Cherepashchuk2018}. 

From the inclination angle we can derive estimates to the mean inclination of the system and the precession angle that the jet  sustain with respect to the axis of rotation. By fitting a linear function to the second half of precessional phases ($>$0.5), we get an inclination of approximately 82 degrees and a precession angle of $\sim$23 degrees which are in complete agreement with the ephemeris of \cite{Eikenberry2001}.

The half opening angle of the jet can range between 1 up to 6 degrees, with lower values (but greater relative errors) on intermediate phases (0.15--0.85). This comes from the fact that on these phases the width of the emission lines becomes harder to constrain as they start to overlap, and with \nustar's resolution they cannot be resolved separately.

An interesting result comes from comparing the expansion velocity of the jet perpendicular to the jet axis ($\beta$~$\sin\Theta$), and the sound speed in the rest frame of the flowing gas 
\begin{equation}
    \beta_s = v_s/c = \sqrt{\frac{5kT_{o}}{3\mu(1+X)m_pc^2}},
\end{equation}
where $kT_{o}$ is the measured temperature of the gas at the base of the jet, m$_p$ is the proton mass, $\mu\sim0.62$ the mean molecular weight and $X=n_i/n_e\sim0.91$ the ion to electron ratio.
We show this relationship on the left panel of \hyperref[fig:betatemp]{Figure~\ref{fig:betatemp}}. We note that within errors, the relationship between these parameters holds true for small angles where $\sin\Theta\sim\Theta$. By looking at \hyperref[fig:jetpars]{Figure~\ref{fig:jetpars}}, we see that this corresponds to observations on intermediate phases (0.15--0.85) where $\Theta<2$~degrees. 
As suggested by \cite{Marshall2002} this relationship might be physical, interpreted as the jet expanding sideways at the sound speed of plasma at its base.

\subsection{Outflow overview}

By following \cite{Khabibullin2016}, we can estimate the evolution of some of the initial conditions at the base of the jet, using the derived fit parameters. Namely, the height from the jet cone apex where it becomes visible to an observer:
\begin{equation}
    r_{o} \sim 1.2\times10^6~\mathrm{cm} \times {L_k}/(\tau_{e0}\beta^3\Theta),
\end{equation}
and the electron density at this radius:
\begin{equation}
    n_{e0} \sim 1.25\times10^{18}~\mathrm{cm}^{-3} \times \tau_{e0}^2 \beta^3 / {L_k}.
\end{equation}

As shown in the right panel of \hyperref[fig:betatemp]{Figure~\ref{fig:betatemp}}, we see that the jet base (also referred as truncation radius) is of the order of 10$^{10}$~cm ($\sim$1.1$\times$10$^{4}$~$r_g$ for $M_{BH}=3M_{\odot}$), ranging from 0.5 to 5 times this value. By taking averaged values of $\Theta$ and $r_o$ we can estimate the size of the jet base $r_{o}\Theta\sim 5.2 \times10^8$cm ($\sim$600~$r_g$).

The electron density at the jet base $n_{e0}$ ranges from 0.2 to $4 \times10^{14}$~cm$^{-3}$. 
We note that these two quantities follow a simplified version of the continuity equation, with $n_{e0}r_{o}^2\sim4\times10^{34}~cm^{-1}$ remaining constant throughout the jet. 

Finally, we can estimate the mass flow through the jet by combining all the above quantities:
\begin{equation}
    \dot{M} = \mu m_{p} (1+X) n_{e0} \pi r_{o}^2 \Theta^2 \beta c.
\end{equation}
We show this result on the right--bottom panel of \hyperref[fig:betatemp]{Figure~\ref{fig:betatemp}}. We see that the mass flow rate ranges between 0.4 up to $3 \times 10^{-6}$~M$_{\odot}$~yr$^{-1}$. 
Assuming a mass of 3~$M_\odot$ for the compact object \citep{Cherepashchuk2018}, we get a maximum of $\sim$20 times the Eddington mass transfer rate.

For reference, \cite{Marshall2002} obtain a value $r_{o}\sim~2\times10^{10}~$cm which lies very well between our estimates. Conversely, they obtain a higher upper limit of $4 \times 10^{15}$~cm$^{-3}$ for the electron density at the jet base, almost ten times our upper limit. This gives $n_{e0} r_{o}^2 \sim 1.7 \times 10^{36}$~cm$^{-1}$, $\lesssim$100 times greater than our estimate. 

By taking the maximum measured 3--70 keV \bjet\ luminosity ($2\times10^{38}$~erg~s$^{-1}$), we can compute the photoionization degree $\xi$ over the spherical region of size $r_o$ and electron density $n_{e0}$, 
$\log~\xi = \log~L_x$/($n_{e0}r_{o}^2) \lesssim 4$. 
This means that the illuminating jet power is sufficient to account for the higher ionization degrees obtained by the reflection components. 
As already stated by \cite{Middleton2021}, the use of more complex reflection models, such as \xcp\ and \rlp, provides a more detailed description of the reflecting medium.


The high absorption column obtained in the spectral fits could be associated with a region of the wind around the jets, which separates the visible part from the invisible one. The density of this region provides an appreciable optical depth for photo-absorption, blocking the jet and thermal disk emission at energies below $\sim$10 keV, but, at the same time, being optically thin for electron scattering; thus partially scattering photons with higher energies.

This concept has already been developed by \cite{Medvedev2018} (\texttt{cwind} model), and they estimate that such condition would require an absorbing column density $N_H$ between $15-20 \times 10^{22}$~cm$^{-2}$ with on optical depth of the order of 0.1. These estimates are in fully in agreement with our obtained values of both parameters.


We can estimate the size of this region if we assume that $n_{e0}R\propto N_H$, ie, a balance between neutral hydrogen and free electrons. By taking average values of $N_H$ and $n_{e0}$ we get $R\sim1.5\times10^9$~cm ($\sim$1700~$r_g$ for $M_{BH}=3M_{\odot}$), which is very similar (within errors and approximations) to the accretion disk spherization radius $R_{sph}\sim1.8\times10^9$~cm$\sim 2000~r_g$ where the accretion regime becomes supercritical \citep{Medvedev2018}. 
This suggests that the absorbing region originates from the combined effect of the high accretion rate, which generates dense gas structures around the compact object in SS~433, and the supercritical disk winds which effectively scatter the soft ($E<10$~keV) photons.


\cite{Middleton2021} attribute the disk wind cone \citep{Dauser2017} as responsible for the lags found at energies up to 9~keV and the hard x-ray excess at 20--30 keV. We frame these results in our scenario by linking the disk wind cone with the combined effect of the reflected spectrum and the central obscuring region.

Specifically, the wind cone model assumes low opening angles ($<10$ degrees) for velocities $\beta\sim$0.2--0.4 to show beaming effects, and a cone height of $10^5~r_g\sim9\times 10^{10}$~cm. Both these model assumptions are in agreement with the values that we found ($\beta\sim$0.28--0.32 and $r_o<10^{11}~$cm).

According to our fitting results, we attribute the reflected spectrum of the accretion disk as responsible for the hard excess component (see \hyperref[fig:fluxes]{Figure~\ref{fig:fluxes}}), and make the case for this obscuring region as the wind itself reprocessing Fe~XXV and Fe~XXVI emission lines (6.7 and 6.97 keV respectively) and thus shifting them onto higher energies.
For this to be possible we followed calculations by \cite{Inoue2022}, who estimated an optical depth of $\sim$1.6 for a compact object of 10~M$_{BH}$ and a radius of 10$^{12}$~cm. If we re-scale by the magnitudes used and obtained in our paper, we find that a lower optical depth of $\tau > 0.02$ is sufficient to account for soft-photon scattering. We find optical depths at the jet base $\tau_{e0}>0.1$  that satisfy the former condition.

\section{Conclusions}
\label{sec:conclusions}

We have reported on the analysis of 10 \nustar\ observations of the Galactic microquasar SS433 that span 1.5 precessional cycles which were taken on almost the same orbital phase. 
We model the averaged spectra with a combination of two precessing thermal jet (\bjet; \citet{Khabibullin2016}) and cold (\xcp; \citet{Garcia2013}) and relativistic reflection (\xcp; \citet{Dauser2014}) emission from an black body type accretion disk (\dbb). We also included Doppler shifting (\zsh) and broadening (\texttt{gsmooth}) components, as well as local and Galactic absorption (\texttt{tbabs}).

Our main results are summarised as follows:
\begin{enumerate}
    \item Jet bulk velocity ranges between 0.28--0.32$c$ and the jet half opening angle is $\lesssim$6~degrees.
    \item The \bjet\ kinetic luminosity ranges between $2-20 \times 10^{38}$~erg~s$^{-1}$, with an average base temperature of $\sim$16~keV and a nickel to iron ratio of $\sim$9.
    \item The western jet relative flux with respect to the eastern jet flux ranges from 0.2 on extreme phases up to $\sim$1 on intermediate phases.
    \item The \dbb\ component gives an inner disk temperature of $\sim$0.38~keV with an inner radius of $\lesssim$30~$r_g$.
    \item The total 3--70 keV luminosity of both jet and disk reflection components range between $0.2-2 \times 10^{38}$~erg~s$^{-1}$, with the jet being completely soft X-ray dominated (3--10 keV), and the disk reflection components hard X-ray dominated (10--70 keV).
    \item We find that at low half opening angles ($\lesssim 2^\circ$), the jet sideways velocity, $\beta\sin\Theta$, can be expressed in terms of the jet base temperature, indicating that it follows an adiabatic expansion regime.
    \item The unabsorbed jet luminosity $L_x\lesssim2\times10^{38}$~erg~s$^{-1}$ is sufficient to account for the high ionization degrees (log~$\xi\lesssim$4) obtained from the reflection components.
    \item The central source and lower parts of the jets could be hidden by an optically thick region of $\tau>0.1$ and size $R\sim N_H/n_{e0}\sim1.5\times10^9$~cm$\sim1700$~$r_g$ for $M_{BH}=3~M_{\odot}$. 
    \end{enumerate}


\begin{acknowledgements}

FAF, JAC and FG acknowledge support by PIP 0113 (CONICET). This work received financial support from PICT-2017-2865 (ANPCyT). FAF is fellow of CONICET. JAC and FG are CONICET researchers. JAC is a Mar\'ia Zambrano researcher fellow funded by the European Union -NextGenerationEU- (UJAR02MZ). 
JAC, JM, FG and PLLE were also supported by grant PID2019-105510GB-C32/AEI/10.13039/501100011033 from the Agencia Estatal de Investigaci\'on of the Spanish Ministerio de Ciencia, Innovaci\'on y Universidades, and by Consejer\'{\i}a de Econom\'{\i}a, Innovaci\'on, Ciencia y Empleo of Junta de Andaluc\'{\i}a as research group FQM-322, as well as FEDER funds.

\end{acknowledgements}

\appendix

\section{}
On \hyperref[tab:modparams]{Table~\ref{tab:modparams}} we present the complete best-fit parameters with errors reported to 90\% confidence level, extracted from MCMC chains of 7$\times$10$^6$ steps (after burning-in the same amount), obtained using 360 walkers (20 times the number of free parameters).

To check for MCMC convergence, we visually inspected the chains of each parameter and determined the most appropriate number of burn-in steps in order to obtain uncorrelated series for the parameters of interest. We corroborated this method by computing the integrated autocorrelation time associated with each series, and verified that it remained as close as unity as possible.

\begin{figure}[h!]
	\centering
	\includegraphics[width=0.5\textwidth]{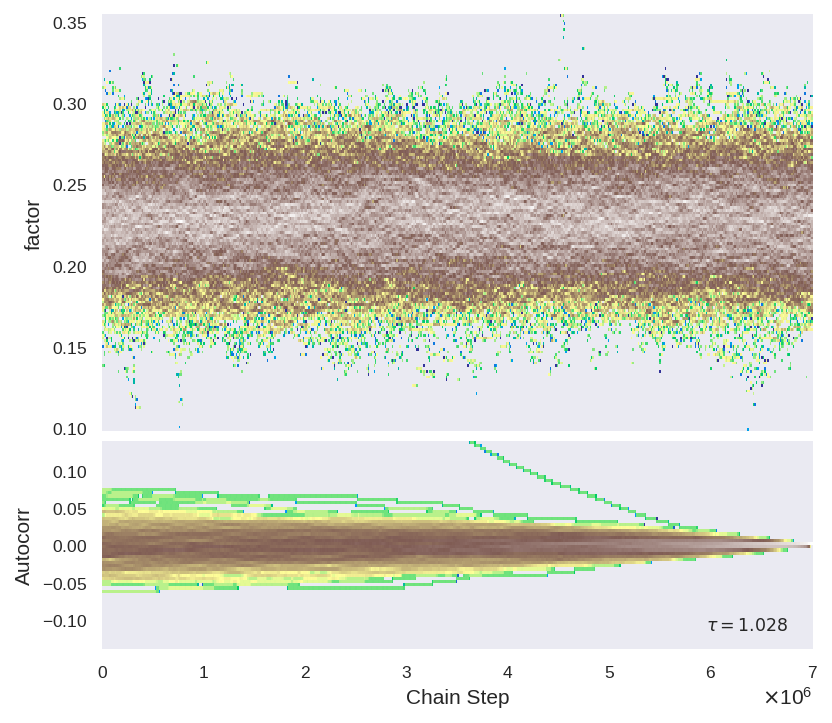}
    \caption{Example of 'trace' plot derived from a MCMC chain. The integrated autocorrelation time $\tau$ is close to unity, which serves as an indicator of the chain convergence. Color gradient indicates density of data points.}
    \label{fig:trace}
\end{figure}

We show an example of a chain 'trace' plot which converged on \hyperref[fig:trace]{Figure~\ref{fig:trace}}. The integrated autocorrelation time $\tau$ is very close to unity, which serves as an numerical indicator of the chain convergence.

\vspace{1cm}

\begin{table*} 
\centering
\renewcommand{\arraystretch}{1.5} 
\begin{tabular*}{\textwidth}{l l l l l l l l l l}
\hline
\hline
\multicolumn{10}{c}{\cons\ $\times$ \tba\ $\times$ \tba\ $\times$ ( \zsh\ $\times$ \bjet\ + \cons\ $\times$ \zsh\ $\times$ \bjet\ + \dbb\ + \xcp\ + \rlp)} \\
\hline
\hspace{5mm}
Obs & $N_H$ & $\Sigma^{\rm jet}_{\rm 6 keV}$ & $Z^{\rm jet}_{\rm east}$ & $L_{38}$ * $\tau_{e0}$ & $T_{o}$ & $A^{\rm jet}_{\rm Fe}$ & $A^{\rm jet}_{\rm Ni}$ & $C^{\rm jet}_{\rm west}$ & $Z^{\rm jet}_{\rm west}$ \\
\hline
02 & $8^{+4}_{-2}$ & $0.03^{+0.06}_{-0.003}$ & $0.008^{+0.003}_{-0.0008}$ & $10^{+4}_{-1}$ & $11^{+0.9}_{-1}$ & $1.1^{+0.1}_{-0.2}$ & $14^{+2}_{-0.9}$ & $0.92^{+0.08}_{-0.06}$ & $0.081^{+0.005}_{-0.001}$ \\
04 & $12^{+1}_{-2}$ & $0.09^{+0.03}_{-0.01}$ & $-0.062^{+0.002}_{-0.001}$ & $25^{+7}_{-1}$ & $14^{+0.5}_{-2}$ & $1.2^{+0.02}_{-0.1}$ & $9^{+0.9}_{-1}$ & $0.32^{+0.02}_{-0.04}$ & $0.15^{+0.004}_{-0.005}$ \\
06 & $14^{+0.7}_{-1}$ & $0.08^{+0.02}_{-0.02}$ & $-0.097^{+0.001}_{-0.0005}$ & $30^{+10}_{-2}$ & $18^{+1}_{-0.9}$ & $2.2^{+0.3}_{-0.5}$ & $15^{+2}_{-3}$ & $0.12^{+0.03}_{-0.03}$ & $0.19^{+0.007}_{-0.01}$ \\
08 & $13.2^{+0.7}_{-0.5}$ & $0.124^{+0.007}_{-0.005}$ & $-0.1059^{+0.0006}_{-0.0005}$ & $21^{+3}_{-2}$ & $16.4^{+0.7}_{-0.6}$ & $2.2^{+0.3}_{-0.3}$ & $13^{+2}_{-2}$ & $0.23^{+0.02}_{-0.02}$ & $0.205^{+0.004}_{-0.005}$ \\
10 & $15^{+2}_{-2}$ & $0^{+0.05}_{-0.002}$ & $-0.083^{+0.001}_{-0.001}$ & $14^{+8}_{-2}$ & $15^{+1}_{-2}$ & $2.4^{+0.3}_{-0.9}$ & $17^{+3}_{-8}$ & $0.19^{+0.06}_{-0.05}$ & $0.19^{+0.01}_{-0.008}$ \\
12 & $14^{+2}_{-3}$ & $0.05^{+0.02}_{-0.04}$ & $0.073^{+0.005}_{-0.002}$ & $8^{+3}_{-0.4}$ & $18^{+3}_{-3}$ & $2.6^{+0.3}_{-0.6}$ & $29^{+1}_{-6}$ & $0.9^{+0.1}_{-0.05}$ & $0.006^{+0.003}_{-0.001}$ \\
14 & $11^{+1}_{-3}$ & $0^{+0.04}_{-0.0005}$ & $0.02^{+0.002}_{-0.002}$ & $11^{+3}_{-1}$ & $13^{+2}_{-1}$ & $2.4^{+0.4}_{-0.3}$ & $21^{+3}_{-2}$ & $0.9^{+0.1}_{-0.1}$ & $0.068^{+0.004}_{-0.001}$ \\
16 & $12^{+1}_{-1}$ & $0.04^{+0.02}_{-0.03}$ & $-0.019^{+0.001}_{-0.002}$ & $17^{+3}_{-2}$ & $16^{+0.8}_{-1}$ & $1.3^{+0.08}_{-0.1}$ & $12^{+1}_{-1}$ & $0.69^{+0.07}_{-0.04}$ & $0.112^{+0.002}_{-0.003}$ \\
18 & $18^{+1}_{-0.8}$ & $0.1^{+0.02}_{-0.01}$ & $-0.093^{+0.0005}_{-0.001}$ & $27^{+4}_{-5}$ & $22^{+0.9}_{-3}$ & $2.3^{+0.3}_{-0.3}$ & $13^{+2}_{-2}$ & $0.08^{+0.04}_{-0.03}$ & $0.2^{+0.009}_{-0.02}$ \\
20 & $13^{+1}_{-2}$ & $0.01^{+0.04}_{-0.007}$ & $0.073^{+0.003}_{-0.001}$ & $17^{+2}_{-3}$ & $15^{+2}_{-1}$ & $1.1^{+0.2}_{-0.08}$ & $13^{+1}_{-0.8}$ & $0.98^{+0.09}_{-0.08}$ & $0.015^{+0.002}_{-0.001}$ \\
\hline
\hspace{5mm}
Obs & $kT_{bb}$ & $N_{bb}$ & $\Gamma_{in}$ & $\log~\xi$ & $\phi$ & $N_{\rm xi}$ & $h$ & $C_{AB}$ & $\chi^2$ / dof \\
\hline
02 & $0.45^{+0.05}_{-0.05}$ & $0.02^{+0.1}_{-0.02}$ & $1.4^{+0.1}_{-0.07}$ & $2.7^{+0.07}_{-0.2}$ & $82^{+2}_{-5}$ & $1.2^{+0.2}_{-0.4}$ & $20^{+70}_{-10}$ & $1.01^{+0.02}_{-0.004}$ & 669.52/642 \\
04 & $0.39^{+0.01}_{-0.01}$ & $1.2^{+0.6}_{-0.4}$ & $1.9^{+0.04}_{-0.1}$ & $2.8^{+0.04}_{-0.2}$ & $70^{+2}_{-3}$ & $9^{+1}_{-2}$ & $2.9^{+0.8}_{-0.5}$ & $1.024^{+0.006}_{-0.005}$ & 1137.00/1016 \\
06 & $0.39^{+0.003}_{-0.02}$ & $2.6^{+0.8}_{-0.4}$ & $1.71^{+0.02}_{-0.04}$ & $3.6^{+0.05}_{-0.1}$ & $60^{+6}_{-10}$ & $12^{+2}_{-2}$ & $2.2^{+0.5}_{-0.2}$ & $1.035^{+0.004}_{-0.005}$ & 1448.34/1333 \\
08 & $0.418^{+0.007}_{-0.004}$ & $1.3^{+0.2}_{-0.1}$ & $1.62^{+0.02}_{-0.02}$ & $3.47^{+0.07}_{-0.07}$ & $50^{+4}_{-4}$ & $8.3^{+0.6}_{-0.9}$ & $3.7^{+0.6}_{-0.7}$ & $1.047^{+0.003}_{-0.003}$ & 1405.17/1340 \\
10 & $0.37^{+0.01}_{-0.02}$ & $1.8^{+0.8}_{-0.6}$ & $1.75^{+0.03}_{-0.05}$ & $3.6^{+0.1}_{-0.2}$ & $20^{+20}_{-20}$ & $1.8^{+0.2}_{-0.5}$ & $10^{+20}_{-3}$ & $1.01^{+0.01}_{-0.009}$ & 686.64/674 \\
12& $0.38^{+0.02}_{-0.009}$ & $1.4^{+0.6}_{-0.7}$ & $1.7^{+0.06}_{-0.09}$ & $3.8^{+0.1}_{-0.2}$ & $10^{+10}_{-8}$ & $1.7^{+0.3}_{-0.2}$ & $9^{+5}_{-2}$ & $1.042^{+0.007}_{-0.007}$ & 851.57/894 \\
14 & $0.42^{+0.01}_{-0.03}$ & $0.3^{+0.1}_{-0.1}$ & $1.7^{+0.05}_{-0.06}$ & $3.7^{+0.1}_{-0.1}$ & $40^{+6}_{-20}$ & $1.4^{+0.2}_{-0.3}$ & $20^{+40}_{-10}$ & $1.03^{+0.003}_{-0.01}$ & 903.51/864 \\
16 & $0.36^{+0.01}_{-0.01}$ & $1.8^{+0.7}_{-0.6}$ & $1.93^{+0.04}_{-0.07}$ & $2.67^{+0.07}_{-0.05}$ & $65^{+3}_{-2}$ & $4^{+1}_{-0.6}$ & $6^{+4}_{-2}$ & $1.055^{+0.006}_{-0.005}$ & 989.56/987 \\
18 & $0.38^{+0.006}_{-0.01}$ & $5^{+2}_{-0.8}$ & $1.74^{+0.02}_{-0.04}$ & $3.6^{+0.04}_{-0.1}$ & $20^{+30}_{-8}$ & $9^{+2}_{-0.9}$ & $2^{+2}_{-0.1}$ & $1.022^{+0.005}_{-0.004}$ & 1383.68/1301 \\
20 & $0.36^{+0.02}_{-0.01}$ & $1.9^{+0.6}_{-0.8}$ & $1.91^{+0.09}_{-0.04}$ & $2.7^{+0.04}_{-0.1}$ & $60^{+6}_{-6}$ & $4^{+1}_{-0.6}$ & $10^{+20}_{-7}$ & $1.015^{+0.004}_{-0.007}$ & 944.15/929 \\
\hline

\end{tabular*}
\caption{
Complete model best fit parameters and fit statistics.\\ \\
$N_H$ : local absorption column density in 10$^{22}$~cm$^{-2}$ units. \\ \\
$\Sigma^{\rm jet}_{\rm 6  keV}$ : Gaussian smoothing factor at $E=6$~keV in eV units. \\ \\
$Z^{\rm jet}_{\rm east}$, $Z^{\rm jet}_{\rm west}$ : eastern and western jet redshifts. \\ \\
$C^{\rm jet}_{\rm west}$ : western jet attenuation factor. \\ \\
$L_{38}$ * $\tau_{e0}$ : jet kinetic luminosity weighted by electron transverse opacity in 10$^{38}$~erg/s units . \\ \\
$T_{o}$ : jet base temperature in keV units. \\ \\
$A^{\rm jet}_{\rm Fe}$, $A^{\rm jet}_{\rm Ni}$ : jet iron and  nickel  abundances in solar units. \\ \\ 
$kT_{\rm bb}$ : \dbb\ temperature in keV units. \\ \\
$N_{\rm bb}$ : \dbb\ normalization ($\times$10$^4$). \\ \\
$\Gamma_{in}$ : \xcp\ incident powerlaw index. \\ \\
$\log~\xi$ : \xcp\ ionization degree. \\ \\
$\phi$ : \xcp\ inclination angle in degree units. \\ \\
$h$ : \rlp\ illuminating source height in gravitational radii units. \\ \\
$N_{\rm xi}$ : \xcp\ (equal to \rlp) normalization  ($\times$10$^{-4}$). \\ \\
$C_{AB}$ : FPMA/B cross correlation factor. }
\label{tab:modparams}
\end{table*}

\begin{figure*}
	\centering
	\includegraphics[width=0.95\textwidth]{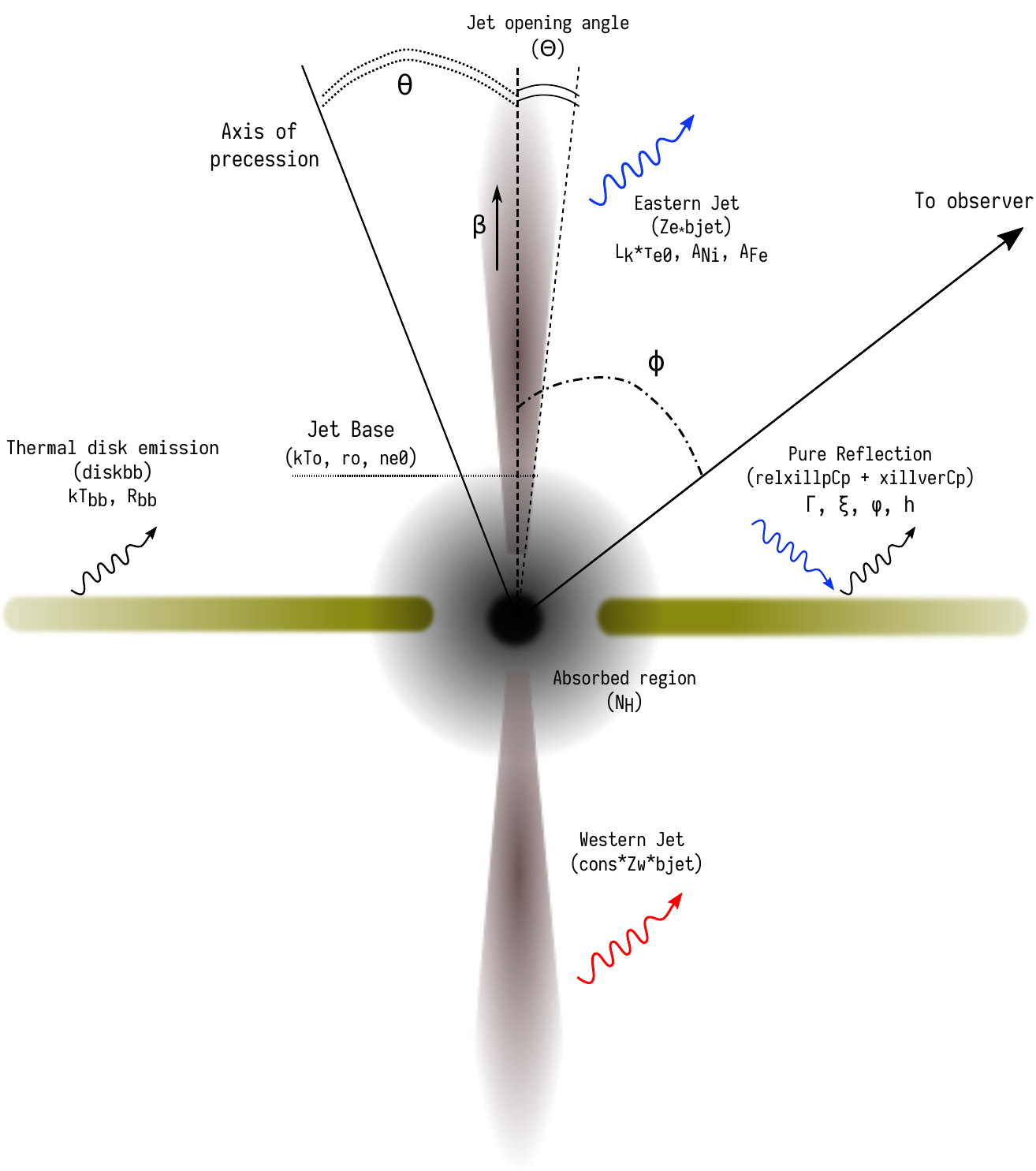}
    \caption{Schematic view of SS433. Each model component is indicated, with the most relevant parameters of the system.}
    \label{fig:squema}
\end{figure*}

On \hyperref[fig:squema]{Figure~\ref{fig:squema}} we present a schematic picture of the microquasar SS433, where the X-ray emission from the jets and the accretion disk components of our scenario are depicted. We also indicate the different geometrical parameters involved, together with specific physical parameters of the model.

\end{document}